\documentclass[aps, prb, twocolumn, superscriptaddress, longbibliography]{revtex4-1}
\usepackage{graphicx}
\usepackage{dcolumn}
\usepackage{bm}
\usepackage{natbib}
\usepackage{upgreek}
\usepackage{amsmath}
\usepackage{amssymb}
\usepackage{hyperref}
\hypersetup{
colorlinks=true,
linkcolor=black,
citecolor=black,
urlcolor=blue
}

\DeclareMathOperator{\realpart}{Re}

\begin{document}

\title{High-frequency breakdown of the integer QHE in GaAs/AlGaAs heterojunctions}

\author{V. Dziom}
\author{A. Shuvaev}
\affiliation{Institute of Solid State Physics, Vienna University of
Technology, 1040 Vienna, Austria}
\author{A. V. Shchepetilnikov}
\affiliation{Institute of Solid State Physics RAS, 142432 Chernogolovka, Moscow district, Russia}
\author{D. MacFarland}
\author{G. Strasser}
\affiliation{Center for Micro- and Nanostructures, Vienna University of
Technology, 1040 Vienna, Austria}
\author{A. Pimenov}
\affiliation{Institute of Solid State Physics, Vienna University of
Technology, 1040 Vienna, Austria}

\begin{abstract}
The integer quantum Hall effect is a well-studied phenomenon at frequencies below about 100\,Hz.
The plateaus in high-frequency Hall conductivity were experimentally proven to retain up to 33\,GHz, but the
behavior at higher frequencies has remained largely unexplored.
Using continuous wave THz spectroscopy, the complex
Hall conductivity of GaAs/AlGaAs heterojunctions was studied in the range of 69--1100\,GHz.
Above 100\,GHz, the quantum plateaus are strongly smeared out and replaced
by weak quantum oscillations in the real part of the conductivity.
The amplitude of the oscillations decreases with increasing frequency.
Near 1\,THz, the Hall conductivity does not reveal any features related
to the filling of Landau levels.
Similar oscillations are observed in the imaginary part as well, this effect has no analogy at zero frequency.
This experimental picture is in disagreement with existing theoretical considerations of the high-frequency quantum Hall effect.
\end{abstract}

\date{\today}

\maketitle

\section{Introduction}
The discovery of the integer quantum Hall effect (IQHE)\cite{Klitzing80} has attracted much interest in scientific community.
The vast majority of experimental and theoretical investigations
have been devoted to the study of the QHE at frequencies below 100\,Hz, and in this range the phenomenon
of Hall quantization has been studied very extensively.
In experiments on QHE it is more convenient to apply an alternating current, rather than a direct current.
At several Hertz $\sigma_{xy}(\omega)$ is indistinguishable from the DC Hall conductivity.
Measurable frequency dependence can be detected when $\omega$ is increased up to the microwave range. In this range the standard contact techniques become inapplicable and high-frequency Hall conductivity is be studied using the interaction of electromagnetic waves with a two-dimensional electron gas (2DEG).
Kuchar \emph{et al.}\cite{Kuchar} used a crossed waveguide setup to observe Hall quantization at 33\,GHz.
Galchenkov \emph{et al.}\cite{Galchenkov70} used a circular waveguide to study the evolution of the Hall plateaus in the 24--70\,GHz range.
A review of experiments on the longitudinal conductivity $\sigma_{xx}(\omega)$ below 20\,GHz can be found in Ref.\,[\onlinecite{Hohls_PRL_2002}].

Further frequency increases can be achieved in quasi-optical spectrometers, suitable for measurements in the range of 100--1000\,GHz.
In the case of a thin conducting film, the Hall conductivity is directly related to the Faraday rotation angle\cite{Volkov}.
Recent experimental works \cite{ShimanoGraphene, OkadaAnomalous, WuBi, DziomQ2759} on the observation of the quantized Faraday rotation in novel materials have inspired a development of theories for a non-linear Hall response \cite{TseNonlinear,LeeNonlinear}.
A linear high-frequency Hall response is far from being
completely understood for systems with parabolic electron bands (AlGaAs, Si, Ge).
Experimentally, the Hall effect in the THz range was observed in GaAs/AlGaAs heterojunctions \cite{IkebeGaAs,StierGaAs}
and in Ge quantum wells \cite{FaillaGe}. The high-frequency data in Refs.\,[\onlinecite{IkebeGaAs,StierGaAs,FaillaGe}]
do not demonstrate quantum plateaus that would be comparable with corresponding DC data. In Ref.\,[\onlinecite{StierGaAs}] the
experiment was conducted at two frequencies (2.52 and 3.14\,THz), using an optically pumped molecular gas laser.
In Refs.\,[\onlinecite{IkebeGaAs},\onlinecite{FaillaGe}] the Hall conductivity was measured using time-domain spectroscopy (TDS).
In principle, TDS allows the Hall conductivity to be obtained at fixed frequencies, but the authors present the data averaged over a wide spectral range. Due to this averaging, information about the frequency dependence of the Hall conductivity was lost.
Thus, the question, of how the static QHE transforms into a dynamic one, remains unresolved.

\begin{figure}[tbp]
\centerline{\includegraphics[width=1.00\linewidth, clip]{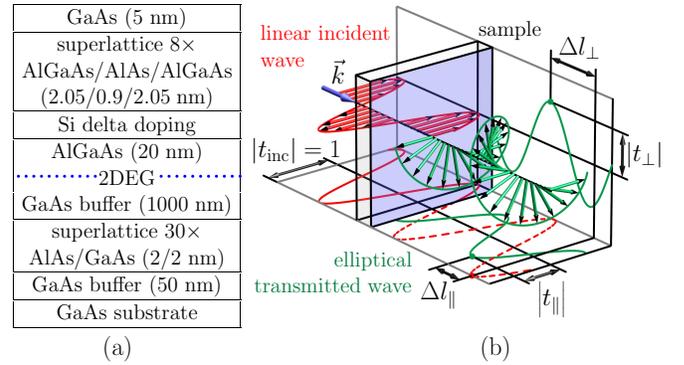}}
\caption{(a) Detailed structure of GaAs/AlGaAs heterojunctions.
The aluminum fraction in Al$_x$Ga$_{1-x}$As is $x = 31.35$\%.
The samples \#1 and \#2 differ by the amount of silicon in the doping layer (Tab.\,\ref{sampleTable}).
(b) Transmission spectroscopy of a two-dimensional electron gas.
The linearly polarized incident wave becomes elliptically polarized upon passing
through the electron gas in a magnetic field.
Using a polarizer (not shown), transmission amplitudes of the linear
components were measured along initial ($t_{\parallel}$)
and perpendicular ($t_{\perp}$) directions. A second reference beam, schematically
shown by the red dashed line, was used to determine the
phase shift $\Delta l$ produced by the sample. The knowledge of  two
complex transmission coefficients
$t_{\parallel}=\left|t_{\parallel} \right| e^{\imath k \Delta l_{\parallel}}$
and
$t_{\perp}=\left|t_{\perp} \right| e^{\imath k \Delta l_{\perp}}$
is sufficient to calculate the high-frequency complex Hall conductivity without
additional model assumptions.
}
\label{fig1}
\end{figure}%

In order to study the evolution of the quantum Hall plateaus with respect to frequency,
a series of experiments were conducted on MBE-grown GaAs/AlGaAs heterojunctions, one of the most suitable system to investigate the DC QHE. The results of the crossed-waveguide
method were reproduced \cite{Kuchar,Galchenkov70} and a QHE plateau was observed at 69\,GHz. Above 100\,GHz the plateaus were replaced by oscillations that disappeared completely as the frequency approached 1\,THz.

\section{Samples and experimental technique}

{
\renewcommand{\arraystretch}{1.4}
\begin{table}
\caption{Parameters of GaAs/AlGaAs heterostructures at $T=1.9$\,K.
$n_{\text{2D}}$: two-dimensional density,
$\mu$: mobility,
$\tau$: relaxation time,
$m$: cyclotron mass,
size: dimensions of the substrate.
The superscripts ``DC'' and ``THz'' denote the quantities,
independently obtained in DC and spectroscopic experiments, respectively.
}
\begin{tabular}{cccc}
\hline
&  Sample \#1 & Sample \#2 \\
\hline \hline
$n_{\text{2D}}^{\text{DC}}$(cm$^{-2}$) & $(2.3\pm0.2)\times 10^{11}$ & $(3.6\pm0.3)\times 10^{11}$ \\
\hline
$n_{\text{2D}}^{\text{THz}}$(cm$^{-2}$) & $(2.4\pm0.2)\times 10^{11}$ & $(3.9\pm0.8)\times 10^{11}$ \\
\hline
$\mu^{\text{DC}}$(cm$^{2}$/(V$\cdot$s)) & $(1.0\pm0.1)\times 10^{5}$ & $(3.2\pm0.5)\times 10^{5}$\\
\hline
$\mu^{\text{THz}}$(cm$^{2}$/(V$\cdot$s)) & $(1.1\pm0.1)\times 10^{5}$ & $(2.5\pm0.5)\times 10^{5}$\\
\hline
$\tau$(ps)&$4.5\pm0.5$&$10\pm2$\\
\hline
$m/m_0$&$0.070\pm0.001$&$0.070\pm0.002$\\
\hline
size (mm$^3$)&10$\times$10$\times$0.660&5$\times$5$\times$0.367\\
\hline
\end{tabular}
\label{sampleTable}
\end{table}
}

The experimental data, presented in this paper, have been obtained on two
GaAs/AlGaAs heterojunctions, grown by molecular beam epitaxy (see Fig.\,\ref{fig1}(a)).
Characteristic parameters of the samples, obtained in DC and spectroscopic experiments at 1.9\,K, are given in Tab.\,\ref{sampleTable}.
Sample \#1 had a reduced silicon delta doping level in comparison with sample \#2, which led to
a lower electron density and a shorter relaxation time.
Insulating GaAs, used as a substrate, was transparent to the radiation in the full range of the spectrometer.
The substrate was characterized by a dielectric constant $\varepsilon=12$ with a negligible frequency dependence.
Indium electrical contacts \cite{SEBESTYEN_SSE_1982,Lakhani_JAP_1984}, placed in the corners and the center of the
sides, were prepared on each sample by baking at 400$^{\circ}$C
in a reducing atmosphere (Ar+4\%H). All spectroscopic experiments
were accompanied by simultaneous measurements of resistances $R_{xx}$ and $R_{xy}$ using
lock-in techniques; typical values of the applied current were $I\approx 1\,\mu$A. The application of the relatively high current without non-linear effects was possible because of large dimensions of the samples in comparison with standard Hall bars, see Tab.\,\ref{sampleTable}.
During the experiments the sample was placed into
a superconducting magnet with optical windows, made of $50\,\mu$m mylar films.
The sample volume was filled with liquid helium and pumped down to maintain
the temperature of the sample at 1.9\,K.

Heterostructures including mixed Al$_x$Ga$_{1-x}$As layers (Fig.\,\ref{fig1}(a)) are known to demonstrate a strong effect of persistent photoconductivity \cite{BabaJJAP83}. For samples cooled in the dark the subsequent illumination with visible light leads to an increase of  the electron density \cite{Nathan_SSE_1986}.
The photo-induced charge carriers persist in the sample even after the switching off of the light and they may create an additional conductive channel parallel to the 2D electrons.
In experiments on illuminated samples \#1 and \#2 the longitudinal DC resistance $R_{xx}$ acquired non-zero values at integer filling factors and the shape of Hall plateaus became distorted after illumination.
In order to avoid the persistent photoconductivity effects the optical windows were constantly
covered by black paper that blocked visible light. All data presented in this work were obtained on samples in the dark.

The high-frequency Hall conductivity of the two-dimensional electron gas was measured in the range of 69-1100\,GHz using a two-beam Mach-Zehnder interferometer. Backward wave oscillators (BWOs) produced a continuous monochromatic wave that was guided in free space using dielectric lenses, metallic mirrors and free-standing wire-grid polarizers. A quasi-parallel incident beam was focused on a sample by a lens with a diameter of 50\,mm and a focal distance of 140\,mm. The transmitted wave passed through a similar lens to restore a quasi-parallel beam. The intensity of the transmitted beam was measured by a 4.2\,K helium cooled Si bolometer.

Upon passing through the sample, the linearly polarized wave became elliptically polarized, see
Fig.\,\ref{fig1}(b). First, a linear component with the same polarization as in the incident wave was filtered by a wire-grid polarizer. The intensity of this component with the sample in the beam divided by the intensity without the sample gave the absolute value of the complex parallel transmission $\left|t_{\parallel}\right|^2$. The phase shift $\Delta l_{\parallel}$ was measured using a reference beam to obtain the complex parallel coefficient as
$t_{\parallel}=\left|t_{\parallel} \right| e^{\imath k \Delta l_{\parallel}}$, where $k=\omega/c$ is the wave vector.
The polarizer was then rotated by $90^{\circ}$ and the procedure was repeated to obtain a complex crossed transmission coefficient $t_{\perp}=\left|t_{\perp} \right| e^{\imath k \Delta l_{\perp}}$.

\begin{figure*}[tbp]
\centerline{\includegraphics[width=0.95\linewidth,clip]{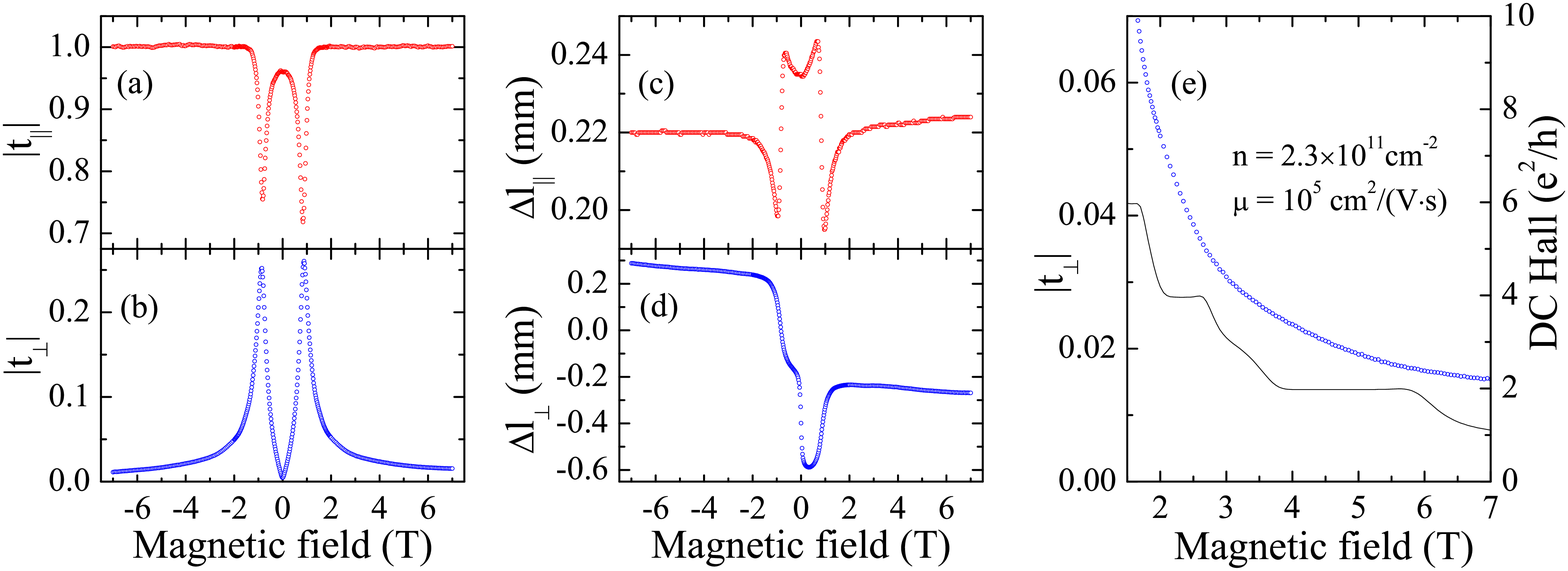}}
\caption{
Field dependence of the transmission coefficients for sample\,\#1 (Tab.\,\ref{sampleTable}) at 332\,GHz. Cyclotron resonance is observed at $\pm0.83$\,T  as a dip in the
parallel amplitude (a) and a peak in the crossed amplitude (b).
The sign change of the external magnetic field does not affect the parallel
transmission ($t_{\parallel}(B)=t_{\parallel}(-B)$), while the complex crossed coefficient changes sign
($t_{\perp}(B)=-t_{\perp}(-B)$). As a result, the parallel phase shift generated by the sample
is an even function of magnetic field (c), and the values of the crossed phase shift differ by half a wavelength (d). The panel (e) shows $\left|t_{\perp}\right|$ near a wide plateau in DC Hall conductance.
}
\label{complexTransmission}
\end{figure*}

In order to obtain the Hall conductivity as a function of the magnetic field, the transmission coefficients were measured at fixed frequencies (Fig.\,\ref{complexTransmission}). The frequency generated by the BWO was controlled by an accelerating voltage and could be set to any value within a certain range. Acting as a Fabry-P\'{e}rot resonator, the dielectric substrate produced regular oscillations in the transmission spectra
(see the upper inset in Fig.\,\ref{figSym}). The frequencies $f_z$, at which the transmission was maximal, were determined by the relation $\sqrt{\varepsilon}k a=\pi z$, where $z$ is an integer. In the framework of a matrix formalism \cite{Berreman}, the substrate is described by a transfer matrix $M$ that connects electromagnetic (EM) fields at the opposite surfaces. At frequencies $f_z$, the transfer matrix of a non-absorbing dielectric slab degenerates
into an identity matrix: $M=(-1)^{z} I$.
At these
frequencies the substrate virtually ``disappears'', as it simply duplicates the EM field at its surfaces.
In transmission coefficients the substrate causes only a phase shift that is equal to the thickness and a sign change, if $z$ is odd.
For sample \#1 ($\varepsilon=12$, $a=0.66$\,mm), the frequencies $f_z$ were multiples of 67\,GHz.
Measuring at one of the transmission maxima allows a higher signal to be obtained, all else being equal.
For this reason, most of the measurements at fixed frequencies $f$,
presented in this work, were carried out at $f\approx f_z$.

The lowest frequency at which the quasi-optical method produces reliable results, is determined by the sample dimensions. The size of the focal spot can be estimated as $f \lambda/D$, where $f=140$\,mm is the focal distance, $D=50$\,mm is the diameter of the lens and $\lambda$ is the radiation wavelength. The measured transmission coefficient starts to be affected by diffraction corrections when the focal spot is comparable with the sample dimensions. Therefore, in case of larger sample \#1 (see Tab.\,\ref{sampleTable}) the spectroscopic measurements could be extended down to frequencies below 100\,GHz.

\section{Data processing}
\begin{figure}
\centerline{\includegraphics[width=1.0\linewidth,clip]{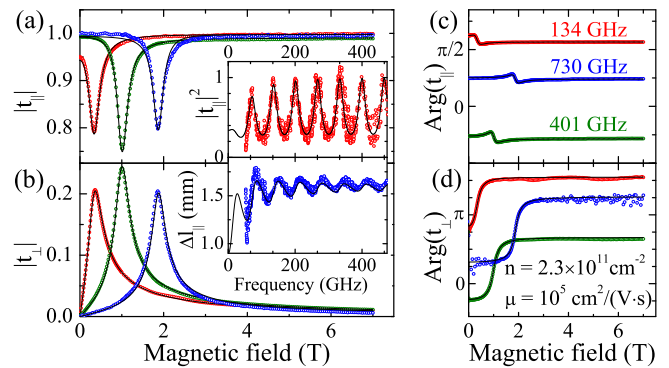}}
\caption{
Symmetrized transmission coefficients for sample\,\#1 at frequencies 134 (red), 401 (green) and
730\,GHz (blue) as a function of
external magnetic field.
The parallel transmission in the zero field is shown as a function of frequency in the inset.
The black solid lines represent classical Drude fits.
}
\label{figSym}
\end{figure}

Knowing the two complex coefficients $t_{\parallel}$ and $t_{\perp}$,
the complex Hall conductivity can be calculated at frequency\,$\omega$ as in Ref.\,[\onlinecite{DziomLight}]:
\begin{equation}
\sigma_{xy}=
\frac
{2 \sqrt{\varepsilon} e^{-\imath k a} t_{\perp}}
{Z_0 (t_{\parallel}^2+t_{\perp}^2)(\sqrt{\varepsilon}\cos{\beta}-\imath\sin{\beta})},
\label{TrToSigma}
\end{equation}
where $a$ is the substrate thickness, $\varepsilon$ is the dielectric constant of the substrate,
$\beta= \sqrt{\varepsilon}k a$ and
$Z_0\approx377$\,$\Omega$ is the impedance of free space.
Equation~(\ref{TrToSigma}) can be simplified in the case
when $\omega$ is close to one of the transmission maxima and
the magnetic field $B$ is much higher than the cyclotron resonance\cite{DziomLight, shuvaev_sst_2012} (CR) field $B_c$.
The vicinity of a maximum corresponds to the value of $\beta=\pi z$, where $z$ is an integer.
If the condition $B\gg B_c$ is satisfied,
then the crossed signal is small and the absorption in 2DEG is negligible: $\left|t_{\perp}\right| \ll \left|t_{\parallel}\right| \simeq 1$,
see Fig.\,\ref{complexTransmission}(a, b). In this case, Eq.\,(\ref{TrToSigma}) can be simplified to
$$
\left|\sigma_{xy}\right|=
\frac
{2}
{Z_0 } \left|t_{\perp}\right|.
$$
Therefore, far from the cyclotron resonance, the plot of directly measured quantity
$\left|t_{\perp}(B)\right|$ represents the absolute value of the Hall conductivity $\left|\sigma_{xy}\right|$, measured in units of $2/Z_0$.
Figure\,\ref{complexTransmission}(e) shows the curve $\left|t_{\perp}(B)\right|$, measured at 332\,GHz, together with the DC Hall conductance.
The $y$-scales in Fig.\,\ref{complexTransmission}(e) are intentionally
mismatched, in order to avoid overlapping data and to clearly demonstrate the absence of any quantum
plateaus in the high-frequency Hall conductivity.

Since the absence of quantization might be a trivial consequence of heating of 2DEG by the radiation \cite{DziomQ2759},
every spectroscopic experiment was accompanied by simultaneous transport measurements. In the vicinity
of QHE plateaus the effect of THz radiation on DC Hall conductance did not exceed 0.3\%. The black solid line
in Fig.\,\ref{complexTransmission}(e) shows DC Hall conductance that was measured simultaneously with
the crossed transmission, shown by blue circles. Thus the experiment demonstrates that the disappearance of
quantum plateaus in high-frequency Hall conductivity is of non-temperature origin. In previous experiments on HgTe quantum wells\cite{DziomQ2759} we checked possible effects of heating by THz radiation. In typical experimental conditions the temperature change was $\sim 0.05$\,K, which supports the arguments above.

The linear polarization of electromagnetic radiation is strictly defined in case of an infinite plane wave only.
In the ideal case,
the complex coefficient $t_{\parallel}(B)$ is an even function of the magnetic field, and $t_{\perp}(B)$ is an odd function.
In the real experimental setup, the beam is restricted by the size of the optical elements and by the
superconducting magnet. These
factors, along with the imperfections of the polarizers,
led to a depolarization of the optical beam, seen as a deviation from the perfect symmetry in the experimental data
(Fig.\,\ref{complexTransmission}).
To reduce these external contributions, a symmetric part of the experimental complex coefficient $t_{\parallel}$ was taken as
$[t_{\parallel}(B)+t_{\parallel}(-B)]/2$ and an antisymmetric part of $t_{\perp}$ as
$[t_{\perp}(B)-t_{\perp}(-B)]/2$.
These corrected transmission coefficients were used in Eq.\,(\ref{TrToSigma}) to calculate
the Hall conductivity.
Figure\,\ref{figSym} shows an example of (anti)symmetrized transmission data together with the classical Drude fitting curves.
The fitting procedure allowed for the effective cyclotron mass $m$, the relaxation time $\tau$, and the electron density
$n_{\text{2D}}^{\text{THz}}$ to be obtained in 2DEG.
Qualitatively, the mass determined the position of the cyclotron resonance, while the combination of the density
and the relaxation time determined its amplitude and width.
The quantity $\mu^{\text{THz}}=e\tau/m$ was compared with the mobility $\mu^{\text{DC}}$,
obtained from the DC measurements of $R_{xx}$. The electron density was another parameter obtained in DC and THz
experiments independently. Both the density and the mobility were found to be in a good agreement, as it can be seen
from Tab\,\ref{sampleTable}.

\section{High-frequency Hall conductivity}

\subsection{Real part of $\sigma_{xy}(\omega)$}
\begin{figure}[tbp]
\centerline{\includegraphics[width=0.99\columnwidth,clip]{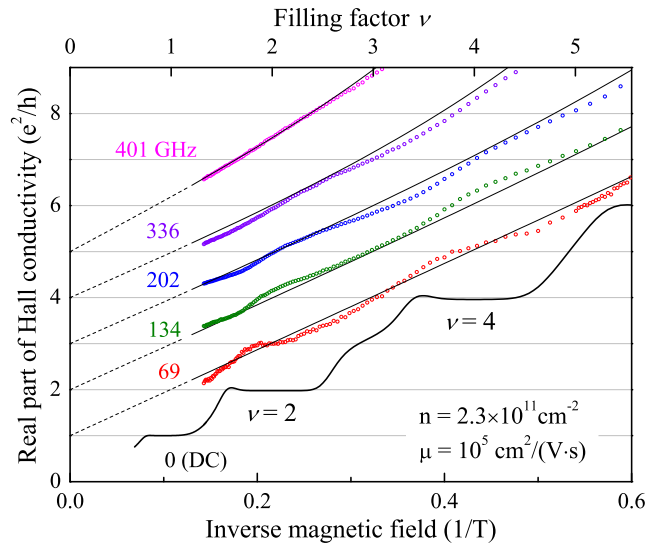}}
\caption{
Evolution of the Hall conductivity with increasing frequency for sample\,\#1 (Tab.\,\ref{sampleTable}).
The DC conductivity, shown by the black solid line, exhibits plateaus
near the even filling factors $\nu = nh/eB$.
High-frequency curves are shifted by 1 for clarity.
At 69\,GHz (red) the real part of $\sigma_{xy}$ has a narrow plateau only around $\nu=2$.
Further increasing of the frequency leads to the smearing of the plateau and to the suppression of
quantum deviations from classical Drude behavior (shown by thin black curves).
}
\label{figMain}
\end{figure}
In order to trace the evolution of the quantum plateaus with increasing frequency, the real part of
the Hall conductivity in sample\,\#1 is plotted as a function of the inverse magnetic field in Fig.\,\ref{figMain}.
The DC conductance, shown by the black curve, demonstrates wide plateaus at even filling factors $\nu = nh/eB$.
In a separate experiment, the DC measurement was extended up to 14\,T. The plateau at $\nu=1$ was also resolved.
The overall behavior of the high-frequency data is well described by the classical Drude theory\cite{Palik}, shown by thin black curves.
At frequencies below 250\,GHz, the cyclotron resonance is located in low magnetic fields, thus the
fitting curves in Fig.\,\ref{figMain} are close to a straight line $\realpart\sigma_{xy}\propto B^{-1}\propto \nu$.
At 69\,GHz (red symbols) a plateau at $\nu=2$ can be detected in the experimental conductivity.
The width of this plateau is 30\% of that in the DC data.
There is no interval with constant $\realpart\sigma_{xy}(B)$ at 134\,GHz (green symbols), and even the slope $\partial \sigma_{xy}/\partial \nu$
does not tend to zero at $\nu=2$. At 134\,GHz, the filling of the second Landau level is revealed as
a slight quantum deviation from the classical curve $\realpart\sigma_{\text{Drude}}(B)$. At higher frequencies, the amplitude of
the quantum deviation decreases. At 401\,GHz (magenta symbols) no signs of the initial plateau can be detected visually on the plot.
The position of the quantum feature in $\realpart\sigma_{xy}(B)$ can be determined by tracking the minimum slope
that shifts to lower magnetic fields with increasing frequency. The plateaus at higher filling factors are smeared out
at 69\,GHz and they disappear in a similar way.

\begin{figure}[tbp]
\centerline{\includegraphics[width=0.95\columnwidth,clip]{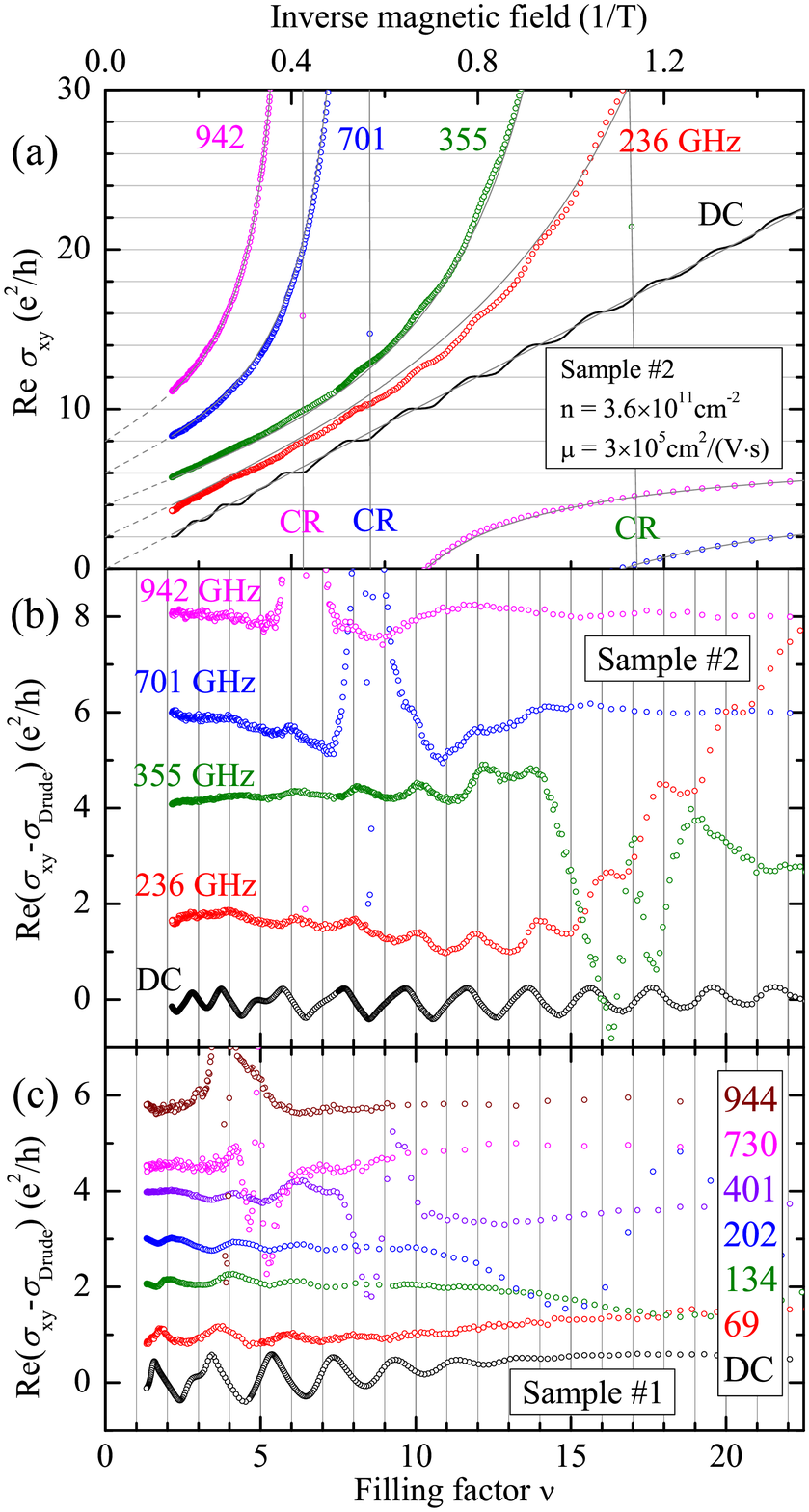}}
\caption{
(a) The Hall conductivity for sample\,\#2 (Tab.\,\ref{sampleTable}) at different frequencies
as a function of the inverse magnetic field and the Landau filling factor $\nu = nh/eB$.
The positions of the cyclotron resonance are denoted by ``CR''.
The solid gray lines are classical Drude fits.
(b, c) The difference between the experimental data and the classical fits for samples\,\#2 and
\#1, respectively. For clarity, the data sets are shifted by $2e^2/h$ in (a, b) and by $e^2/h$ in (c).
All panels are matched by $\nu$ and the magnetic field is indicated for sample\,\#2 (a, b).
}
\label{figRe12}
\end{figure}

In comparison with sample\,\#1, sample\,\#2 had a higher electron mobility and
electron density (see Tab.\,\ref{sampleTable}).
The evolution of the real part of $\sigma_{xy}$ for sample\,\#2 is shown in Fig.\,\ref{figRe12}(a).
Similar to sample\,\#1, the cyclotron resonance in high-frequency Hall conductivity can be approximated by classical
Drude fits. Quantum oscillations, corresponding to the filling of Landau levels,
can be detected in the high-field region as well.
The difference between the experimental conductivity and the classical Drude fit is plotted in Fig.\,\ref{figRe12}(b).
The large discrepancy at $B\approx B_c$ is due to the large value of optical conductivity (approaching 100\,e$^2$/h)
along with the steep slope $\partial \sigma_{xy}/\partial \nu$.
A maximal amplitude of the quantum deviations in the sample\,\#2 is achieved at filling factors of $\nu>10$,
while the oscillations attenuate with increasing filling factor $\nu$ in sample\,\#1, see Fig.\,\ref{figRe12}(c).

Although no flat plateaus can be detected in the high-frequency $\sigma_{xy}$,
the amplitude of the quantum deviations at 236\,GHz (Fig.\,\ref{figRe12}(b)) above $\nu =10$ is comparable with the amplitude of
quantum deviations in the DC conductance.
The phase of the AC deviations is shifted with respect to the DC data.
This effect is better seen in Fig.\,\ref{figRe12}(b) when comparing DC and 236\,GHz curves around $\nu=12$.
The DC $\sigma_{xy}$ and $\sigma_{\text{Drude}}$ coincide at integer filling factors $\nu=10$, 12, 14, \dots,
and their difference $(\sigma_{xy}-\sigma_{\text{Drude}})$ peaks at half integers above and below
(e.g. positive peaks at $\nu=9.5$, 11.5, 13.5, \dots and negative peaks at 10.5, 12.5, 14.5, \dots)
However for the AC data at 236\,GHz the difference has positive peaks at $\nu=10$, 12, 14, \textit{etc}.
A similar phase
shift can be observed in sample\,\#1 around $\nu=3$, see Fig.\,\ref{figRe12}(c). The shift of oscillations in the difference $(\sigma_{xy}-\sigma_{\text{Drude}})$ corresponds to a shift of positions of the minimal slope $\realpart (\partial \sigma_{xy}/\partial \nu)$ towards higher values of $\nu$ and towards smaller magnetic fields.
Similar effect was reported previously \cite{Galchenkov70, IkebeGaAs} and attributed to the difference in the Landau level broadening and in the localization length between adjacent Landau levels.

According to the relation $\Omega_c\propto B \propto 1/\nu_c$, the cyclotron resonance shifts to lower $\nu$ as
the radiation frequency increases. Figures\,\ref{figRe12}(b, c) demonstrate that
the quantum oscillations become attenuated to the right of the CR ($\nu>\nu_c$),
where the radiation frequency exceeds the cyclotron gap ($\omega>\Omega_c$).

\subsection{Imaginary part of $\sigma_{xy}(\omega)$}
\begin{figure*}[tbp]
\centerline{\includegraphics[width=1.00\linewidth,clip]{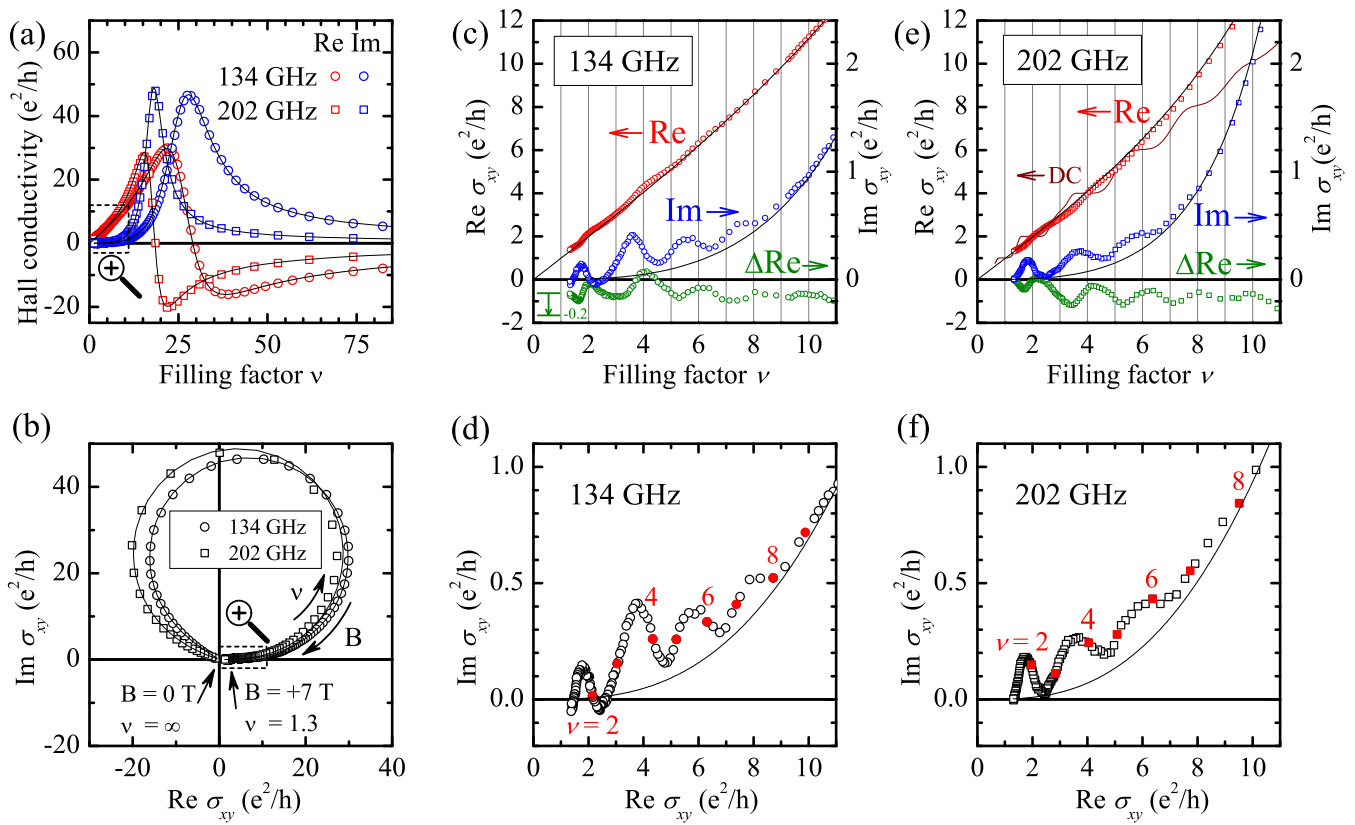}}
\caption{
The complex Hall conductivity $\sigma_{xy}$ in sample\,\#1 (Tab.\,\ref{sampleTable}) at 134 and 202\,GHz.
(a) Real and imaginary parts of $\sigma_{xy}$ are shown as a function of the Landau filing factor $\nu = nh/eB$.
The overall resonance behavior is well described by the Drude model (black solid lines).
Panel (b) shows $\sigma_{xy}$ on a complex plane as a parametric plot with the filling factor as a parameter.
In the low-$\nu$ limit (c--f), where QHE is observed in DC, the imaginary part of $\sigma_{xy}$ demonstrates
substantial periodic deviations from the classical curve (c, e). These deviations are comparable to the deviations in the
real part, shown by green symbols on the same scale. In panels (d, f) the integer values of $\nu$ are indicated by red symbols.
}
\label{figImaginary}
\end{figure*}
While the Hall conductivity is a real number in the static case,
it becomes a complex number with a nonzero imaginary part at finite frequencies.
Figures\,\ref{figImaginary}(a,c,e) show the experimentally obtained $\sigma_{xy}(\nu)$
(symbols) at 134 and 202\,GHz together with the Drude fits (solid curves) for sample\,\#1.
Figures\,\ref{figImaginary}(b,d,f) show $\sigma_{xy}$ on a complex plane as a parametric plot
with the filling factor $\nu$ as a parameter.
The sweep of the magnetic field from 7 to 0\,T corresponds to the change of $\nu$ from 1.32 to $\infty$.
In this representation, classical theory produces a circle-like curve, depicted in Fig.\,\ref{figImaginary}(b)
by black solid lines.
For higher frequencies the curve is getting closer to a perfect circle that is centered on the imaginary axis and
passes through the origin of the coordinates.
The resonance behavior of experimental conductivity is well described by the classical Drude theory,
see Fig.\,\ref{figImaginary}(a, b).
However, when only a few Landau levels are occupied, $\sigma_{xy}$ demonstrates substantial
deviations from the classical curve, see Fig.\,\ref{figImaginary}(c--f).
Due to experimental limitations, the complex argument of $\sigma_{xy}(\nu)$ was determined up to
an unknown constant value, which can be estimated
by comparing with the Drude fit. In Fig.\,\ref{figImaginary} this value was chosen to match the theoretical
and experimental curves near $\nu=10$, where the quantum deviations are faded out.
In this case, the imaginary part of the quantum corrections appears to be positive nearly everywhere and
the imaginary part of $\sigma_{xy}$ tends to preserve its original sign.
As discussed above, in the real part of $\sigma_{xy}$, the deviations can be regarded as remnants of the
DC Hall plateaus. Figures\,\ref{figImaginary}(c, e) show the real part of
the difference $(\sigma_{xy}-\sigma_{\text{Drude}})$, depicted by green symbols on the same scale as the imaginary part.
These plots demonstrate that the quantum oscillations
have similar amplitudes in the real and imaginary parts and that the phases are shifted by $\approx\pi/2$.
The broken periodicity below $\nu=2$ is likely due to the presence of the quantum plateau at
$\nu=1$, which is the only odd plateau resolved in this sample (Fig.\,\ref{figMain}).

\section{Discussion}

The most striking feature of the QHE at zero frequency is the exact quantization of the Hall resistance $R_{xy}$,
which is a macroscopic property of a whole sample, directly obtained in DC experiments. This fact alone does not prove
that $\sigma_{xy}$ is also exactly quantized \cite{KucharScripta}, because local inhomogeneities of the
two-dimensional gas are always present in a real sample.
Unlike the contact techniques, the spectroscopic experiments test $\sigma_{xy}$ directly. Experiments at 30\,GHz
\cite{KucharScripta} demonstrated that the plateaus of non-zero width are also present in $\sigma_{xy}$.
As shown above, the plateaus in $\sigma_{xy}$ disappear at higher frequencies. For the samples in this study
the critical frequency lies near 100\,GHz. Above this frequency the two-dimensional electron gas loses
its QHE features and the Hall conductivity follows the classical Drude model.

Although the IQHE has been extensively studied theoretically, only a few works have
addressed the Hall conductivity in a high-frequency regime \cite{MorimotoPRL09}.
When calculating the Hall conductivity in a linear approximation,
a common approach is to apply linear perturbation theory (Kubo formalism) to a model system.
The theoretical models of IQHE consider non-interacting fermions in a strong magnetic field,
placed in a model potential, which simulates the presence of impurities and constraints in a sample. Depending on
the chosen potential, the analysis of such models can be conducted analytically or numerically.

In Refs.\,[\onlinecite{MorimotoPRL09},\onlinecite{MorimotoPRB10}] the high-frequency Hall conductivity was calculated using a
numerical method of exact diagonalization. In order to model the disorder, the authors treated
randomly distributed Gaussian scatterers.
As calculated within this model, $\sigma_{xy}(\omega)$ was found to retain the Hall plateaus in the THz range.
In Ref.\,[\onlinecite{IkebeGaAs}], these model results were referred to justify the procedure
of averaging $\sigma_{xy}(\omega,\nu)$ over a range of frequencies from 0.5 to 1.2\,THz. The resulting
averaged $\widetilde{\sigma}_{xy}(\nu)$ had a plateau-like feature of vanishing width in comparison with a wide plateau in DC.
This experimental fact, reported in Ref.\,[\onlinecite{IkebeGaAs}], indicates that the plateaus actually smeared out below 1.2\,THz.
The procedure of averaging should be reconsidered, since it only masks the disappearance of the Hall plateaus
for increasing frequencies.

In earlier works, the high-frequency Hall conductivity was treated analytically in two opposite limits: for
scatterers with $\delta$-potential \cite{LozovikDelta} and for a slowly varying potential of impurities \cite{ApenkoSmooth}.
In Ref.\,[\onlinecite{LozovikDelta}], the Hall conductivity was obtained within the $\delta$-impurity model \cite{PrangeDelta}
as a function of electron density $n$. At finite frequencies, the dependence $\realpart\sigma_{xy}(n)$ is predicted to have
a single-dip or
a double-dip structure instead of a flat plateau at DC. A monotonic dependence $\realpart\sigma_{xy}(n)$ is achieved only
if both negative and positive $\delta$-impurities are present in the calculation and the Landau level broadening
exceeds the
cyclotron energy. The last condition was likely not fulfilled in this study’s samples, while the experimental high-frequency
$\realpart\sigma_{xy}(B)$
was monotonic in the vicinity of the DC plateaus.
Unfortunately, the imaginary part of $\sigma_{xy}$ was not treated in Ref.\,[\onlinecite{LozovikDelta}] and
no explicit estimation was given for the critical frequency $f_0$ at which the plateaus were destroyed. However, the consideration can be
extended. In particular, the critical frequency $f_0$ can be shown to be close
to the half-width $\Gamma/(4\pi\hbar)$ of the corresponding broadened Landau level \cite{LozovikPrivateCommunications}.
If the level broadening is assumed to be caused by the scattering on short-range ionized impurities, then the width
can be estimated as \cite{AndoJPs}:
\begin{equation}
\Gamma=\hbar\sqrt{\frac{2\Omega_c}{\pi\tau}}.
\label{LandauBroadening}
\end{equation}

Since the cyclotron frequency $\Omega_c=e B/m$ increases with magnetic field, the plateaus at small filling factors are expected
to be retained at higher radiation frequencies.
Using the parameters in Tab.\,\ref{sampleTable} for sample\,\#1, the critical frequency is obtained as
$f_0=47\sqrt{B}$\,GHz.
For the plateaus at $\nu=2$ and $\nu=4$, the estimated critical frequencies are 102 and 72\,GHz, respectively. In agreement
with this estimation, the plateau at $\nu=2$ at 69\,GHz was experimentally observed. The plateau at $\nu=4$ was not resolved,
as the corresponding critical frequency of 72\,GHz is close to the radiation frequency. The plateaus at $\nu>4$ are already
absent at $f\ge69$\,GHz, since they occur in $R_{xy}$ at even lower magnetic fields.
Due to the longer relaxation time in sample\,\#2 (Tab.\,\ref{sampleTable}), the estimated Landau level width
was smaller and the critical frequency was lower than in sample\,\#1. For this reason, no plateaus in high-frequency conductivity
could be observed in sample\,\#2.

If Eq.\,(\ref{LandauBroadening}) is applied to the case of
HgTe films \cite{ShuvaevUniversal} and to graphene \cite{ShimanoGraphene}, critical frequencies of 1\,THz
and $\approx 3$\,THz are obtained, respectively.
These higher values are formally achieved due to the smaller effective masses and the shorter relaxation times in
the CdHgTe wells and in graphene.
The direct application of Eq.\,(\ref{LandauBroadening}) to the systems with a strongly non-parabolic
dispersion is questionable. However, it is possible that the observation of the quantized Faraday rotation in these materials at
higher frequencies is indeed connected to the larger width of the Landau levels.

In Ref.\,[\onlinecite{ApenkoSmooth}] the Hall conductivity was calculated using the drift
approximation \cite{IordanskySmooth,KazarinovSmooth}. In the limit of very high frequencies, $\sigma_{xy}(\nu)$ tended
to the classical straight line with a small quantum correction:
\[
\sigma_{xy}(\nu)=\nu e^2/h+\delta\sigma_{xy}(\nu).
\]
As calculated within this model, the term $\delta\sigma_{xy}$ has zero imaginary part, while the experimental $\delta\sigma_{xy}$
has both real and imaginary parts of a similar amplitude. The calculation of intermediate frequencies resulted in $\sigma_{xy}$ with
a non-zero imaginary part. However, in this case the dependence of $\realpart \sigma_{xy}(\nu)$ was not monotonic.
Thus, the shape of the smeared quantum plateaus is not described by this approach even qualitatively. The critical frequency calculated in
the drift approximation was connected to the level broadening, similar to the case of $\delta$-potential treated above.

None of the existing theoretical models provides a satisfactory description for the shape of the plateaus in the high-frequency
Hall conductivity. The numerical method of exact diagonalization predicts the persistence of the plateaus in the THz range.
Within this model, a decrease in
disorder leads to a decreasing width $\Gamma$ of Landau levels and to a more distinct quantization in $\sigma_{xy}$. In contrast, the analytical methods predict the destruction of the plateaus at frequencies $f>\Gamma/h$. Within these models, a stronger disorder
allows the quantum plateaus to occur at higher frequencies. The experimental data presented in this work and in
Refs.\,[\onlinecite{KucharScripta,StierGaAs,ShimanoGraphene,ShuvaevUniversal}]
support
the result of the analytical methods. An ultimate understanding requires further systematic experimental studies on samples with significantly
different electron mobilities and cyclotron masses.

\section{Conclusions}

The dynamic quantum Hall effect was studied using continuous-wave THz spectroscopy in the frequency range of 69--1100\,GHz. A clear frequency dependence of the
quantum deviations from the classical Drude model was observed. The extinction of the QHE plateaus took place near
100\,GHz.
Only small quantum corrections were observed above this frequency. Some theoretical models describe this phenomenon
qualitatively, while other models predict the persistence of the plateaus in the high-frequency range. The results
of this work will stimulate efforts towards a complete understanding of the IQHE.

\section{Acknowledgments}

We acknowledge valuable discussion with Yu. Lozovik and Zh. Gevorkyan. This
work was supported by Austrian Science Funds (W--1243, P27098--N27, I3456--N27).

\bibliography{literature}

\begin{thebibliography}{33}%
\makeatletter
\providecommand \@ifxundefined [1]{%
 \@ifx{#1\undefined}
}%
\providecommand \@ifnum [1]{%
 \ifnum #1\expandafter \@firstoftwo
 \else \expandafter \@secondoftwo
 \fi
}%
\providecommand \@ifx [1]{%
 \ifx #1\expandafter \@firstoftwo
 \else \expandafter \@secondoftwo
 \fi
}%
\providecommand \natexlab [1]{#1}%
\providecommand \enquote  [1]{``#1''}%
\providecommand \bibnamefont  [1]{#1}%
\providecommand \bibfnamefont [1]{#1}%
\providecommand \citenamefont [1]{#1}%
\providecommand \href@noop [0]{\@secondoftwo}%
\providecommand \href [0]{\begingroup \@sanitize@url \@href}%
\providecommand \@href[1]{\@@startlink{#1}\@@href}%
\providecommand \@@href[1]{\endgroup#1\@@endlink}%
\providecommand \@sanitize@url [0]{\catcode `\\12\catcode `\$12\catcode
  `\&12\catcode `\#12\catcode `\^12\catcode `\_12\catcode `\%12\relax}%
\providecommand \@@startlink[1]{}%
\providecommand \@@endlink[0]{}%
\providecommand \url  [0]{\begingroup\@sanitize@url \@url }%
\providecommand \@url [1]{\endgroup\@href {#1}{\urlprefix }}%
\providecommand \urlprefix  [0]{URL }%
\providecommand \Eprint [0]{\href }%
\providecommand \doibase [0]{http://dx.doi.org/}%
\providecommand \selectlanguage [0]{\@gobble}%
\providecommand \bibinfo  [0]{\@secondoftwo}%
\providecommand \bibfield  [0]{\@secondoftwo}%
\providecommand \translation [1]{[#1]}%
\providecommand \BibitemOpen [0]{}%
\providecommand \bibitemStop [0]{}%
\providecommand \bibitemNoStop [0]{.\EOS\space}%
\providecommand \EOS [0]{\spacefactor3000\relax}%
\providecommand \BibitemShut  [1]{\csname bibitem#1\endcsname}%
\let\auto@bib@innerbib\@empty
\bibitem [{\citenamefont {Klitzing}\ \emph {et~al.}(1980)\citenamefont
  {Klitzing}, \citenamefont {Dorda},\ and\ \citenamefont
  {Pepper}}]{Klitzing80}%
  \BibitemOpen
  \bibfield  {author} {\bibinfo {author} {\bibfnamefont {K.~v.}\ \bibnamefont
  {Klitzing}}, \bibinfo {author} {\bibfnamefont {G.}~\bibnamefont {Dorda}}, \
  and\ \bibinfo {author} {\bibfnamefont {M.}~\bibnamefont {Pepper}},\
  }\bibfield  {title} {\enquote {\bibinfo {title} {New method for high-accuracy
  determination of the fine-structure constant based on quantized {H}all
  resistance},}\ }\href {\doibase 10.1103/PhysRevLett.45.494} {\bibfield
  {journal} {\bibinfo  {journal} {Phys. Rev. Lett.}\ }\textbf {\bibinfo
  {volume} {45}},\ \bibinfo {pages} {494--497} (\bibinfo {year}
  {1980})}\BibitemShut {NoStop}%
\bibitem [{\citenamefont {Kuchar}\ \emph {et~al.}(1986)\citenamefont {Kuchar},
  \citenamefont {Meisels}, \citenamefont {Weimann},\ and\ \citenamefont
  {Schlapp}}]{Kuchar}%
  \BibitemOpen
  \bibfield  {author} {\bibinfo {author} {\bibfnamefont {F.}~\bibnamefont
  {Kuchar}}, \bibinfo {author} {\bibfnamefont {R.}~\bibnamefont {Meisels}},
  \bibinfo {author} {\bibfnamefont {G.}~\bibnamefont {Weimann}}, \ and\
  \bibinfo {author} {\bibfnamefont {W.}~\bibnamefont {Schlapp}},\ }\bibfield
  {title} {\enquote {\bibinfo {title} {Microwave {H}all conductivity of the
  two-dimensional electron gas in {GaAs}-{Al}$_x${Ga}$_{1-x}${As}},}\ }\href
  {\doibase 10.1103/PhysRevB.33.2965} {\bibfield  {journal} {\bibinfo
  {journal} {Phys. Rev. B}\ }\textbf {\bibinfo {volume} {33}},\ \bibinfo
  {pages} {2965--2967} (\bibinfo {year} {1986})}\BibitemShut {NoStop}%
\bibitem [{\citenamefont {Galchenkov}\ \emph {et~al.}(1987)\citenamefont
  {Galchenkov}, \citenamefont {Grodnenskii}, \citenamefont {Kostovetskii},\
  and\ \citenamefont {Matov}}]{Galchenkov70}%
  \BibitemOpen
  \bibfield  {author} {\bibinfo {author} {\bibfnamefont {L.~A.}\ \bibnamefont
  {Galchenkov}}, \bibinfo {author} {\bibfnamefont {I.~M.}\ \bibnamefont
  {Grodnenskii}}, \bibinfo {author} {\bibfnamefont {M.~V.}\ \bibnamefont
  {Kostovetskii}}, \ and\ \bibinfo {author} {\bibfnamefont {O.~R.}\
  \bibnamefont {Matov}},\ }\bibfield  {title} {\enquote {\bibinfo {title}
  {Frequency dependence of the {H}all conductivity of a {2D} electron gas},}\
  }\href {http://www.jetpletters.ac.ru/ps/1234/article_18636.shtml} {\bibfield
  {journal} {\bibinfo  {journal} {JETP Lett.}\ }\textbf {\bibinfo {volume}
  {46}},\ \bibinfo {pages} {542} (\bibinfo {year} {1987})}\BibitemShut
  {NoStop}%
\bibitem [{\citenamefont {Hohls}\ \emph {et~al.}(2002)\citenamefont {Hohls},
  \citenamefont {Zeitler}, \citenamefont {Haug}, \citenamefont {Meisels},
  \citenamefont {Dybko},\ and\ \citenamefont {Kuchar}}]{Hohls_PRL_2002}%
  \BibitemOpen
  \bibfield  {author} {\bibinfo {author} {\bibfnamefont {F.}~\bibnamefont
  {Hohls}}, \bibinfo {author} {\bibfnamefont {U.}~\bibnamefont {Zeitler}},
  \bibinfo {author} {\bibfnamefont {R.~J.}\ \bibnamefont {Haug}}, \bibinfo
  {author} {\bibfnamefont {R.}~\bibnamefont {Meisels}}, \bibinfo {author}
  {\bibfnamefont {K.}~\bibnamefont {Dybko}}, \ and\ \bibinfo {author}
  {\bibfnamefont {F.}~\bibnamefont {Kuchar}},\ }\bibfield  {title} {\enquote
  {\bibinfo {title} {Dynamical scaling of the quantum {H}all plateau
  transition},}\ }\href {\doibase 10.1103/PhysRevLett.89.276801} {\bibfield
  {journal} {\bibinfo  {journal} {Phys. Rev. Lett.}\ }\textbf {\bibinfo
  {volume} {89}},\ \bibinfo {pages} {276801} (\bibinfo {year}
  {2002})}\BibitemShut {NoStop}%
\bibitem [{\citenamefont {Volkov}\ and\ \citenamefont
  {Mikhailov}(1985)}]{Volkov}%
  \BibitemOpen
  \bibfield  {author} {\bibinfo {author} {\bibfnamefont {V.~A.}\ \bibnamefont
  {Volkov}}\ and\ \bibinfo {author} {\bibfnamefont {S.~A.}\ \bibnamefont
  {Mikhailov}},\ }\bibfield  {title} {\enquote {\bibinfo {title} {Quantization
  of the {F}araday effect in systems with a quantum {H}all effect},}\ }\href
  {http://jetpletters.ac.ru/ps/1467/article_22375.shtml} {\bibfield  {journal}
  {\bibinfo  {journal} {JETP Lett.}\ }\textbf {\bibinfo {volume} {41}},\
  \bibinfo {pages} {476} (\bibinfo {year} {1985})}\BibitemShut {NoStop}%
\bibitem [{\citenamefont {Shimano}\ \emph {et~al.}(2013)\citenamefont
  {Shimano}, \citenamefont {Yumoto}, \citenamefont {Yoo}, \citenamefont
  {Matsunaga}, \citenamefont {Tanabe}, \citenamefont {Hibino}, \citenamefont
  {Morimoto},\ and\ \citenamefont {Aoki}}]{ShimanoGraphene}%
  \BibitemOpen
  \bibfield  {author} {\bibinfo {author} {\bibfnamefont {R.}~\bibnamefont
  {Shimano}}, \bibinfo {author} {\bibfnamefont {G.}~\bibnamefont {Yumoto}},
  \bibinfo {author} {\bibfnamefont {J.~Y.}\ \bibnamefont {Yoo}}, \bibinfo
  {author} {\bibfnamefont {R.}~\bibnamefont {Matsunaga}}, \bibinfo {author}
  {\bibfnamefont {S.}~\bibnamefont {Tanabe}}, \bibinfo {author} {\bibfnamefont
  {H.}~\bibnamefont {Hibino}}, \bibinfo {author} {\bibfnamefont
  {T.}~\bibnamefont {Morimoto}}, \ and\ \bibinfo {author} {\bibfnamefont
  {H.}~\bibnamefont {Aoki}},\ }\bibfield  {title} {\enquote {\bibinfo {title}
  {Quantum {F}araday and {K}err rotations in graphene},}\ }\href
  {http://dx.doi.org/10.1038/ncomms2866} {\bibfield  {journal} {\bibinfo
  {journal} {Nat. Commun.}\ }\textbf {\bibinfo {volume} {4}},\ \bibinfo {pages}
  {1841} (\bibinfo {year} {2013})}\BibitemShut {NoStop}%
\bibitem [{\citenamefont {Okada}\ \emph {et~al.}(2016)\citenamefont {Okada},
  \citenamefont {Takahashi}, \citenamefont {Mogi}, \citenamefont {Yoshimi},
  \citenamefont {Tsukazaki}, \citenamefont {Takahashi}, \citenamefont {Ogawa},
  \citenamefont {Kawasaki},\ and\ \citenamefont {Tokura}}]{OkadaAnomalous}%
  \BibitemOpen
  \bibfield  {author} {\bibinfo {author} {\bibfnamefont {K.~N.}\ \bibnamefont
  {Okada}}, \bibinfo {author} {\bibfnamefont {Y.}~\bibnamefont {Takahashi}},
  \bibinfo {author} {\bibfnamefont {M.}~\bibnamefont {Mogi}}, \bibinfo {author}
  {\bibfnamefont {R.}~\bibnamefont {Yoshimi}}, \bibinfo {author} {\bibfnamefont
  {A.}~\bibnamefont {Tsukazaki}}, \bibinfo {author} {\bibfnamefont {K.~S.}\
  \bibnamefont {Takahashi}}, \bibinfo {author} {\bibfnamefont {N.}~\bibnamefont
  {Ogawa}}, \bibinfo {author} {\bibfnamefont {M.}~\bibnamefont {Kawasaki}}, \
  and\ \bibinfo {author} {\bibfnamefont {Y.}~\bibnamefont {Tokura}},\
  }\bibfield  {title} {\enquote {\bibinfo {title} {Terahertz spectroscopy on
  {F}araday and {K}err rotations in a quantum anomalous {H}all state},}\ }\href
  {http://dx.doi.org/10.1038/ncomms12245} {\bibfield  {journal} {\bibinfo
  {journal} {Nat. Commun.}\ }\textbf {\bibinfo {volume} {7}},\ \bibinfo {pages}
  {12245} (\bibinfo {year} {2016})}\BibitemShut {NoStop}%
\bibitem [{\citenamefont {Wu}\ \emph {et~al.}(2016)\citenamefont {Wu},
  \citenamefont {Salehi}, \citenamefont {Koirala}, \citenamefont {Moon},
  \citenamefont {Oh},\ and\ \citenamefont {Armitage}}]{WuBi}%
  \BibitemOpen
  \bibfield  {author} {\bibinfo {author} {\bibfnamefont {Liang}\ \bibnamefont
  {Wu}}, \bibinfo {author} {\bibfnamefont {M.}~\bibnamefont {Salehi}}, \bibinfo
  {author} {\bibfnamefont {N.}~\bibnamefont {Koirala}}, \bibinfo {author}
  {\bibfnamefont {J.}~\bibnamefont {Moon}}, \bibinfo {author} {\bibfnamefont
  {S.}~\bibnamefont {Oh}}, \ and\ \bibinfo {author} {\bibfnamefont {N.~P.}\
  \bibnamefont {Armitage}},\ }\bibfield  {title} {\enquote {\bibinfo {title}
  {Quantized {F}araday and {K}err rotation and axion electrodynamics of the
  surface states of three-dimensional topological insulators},}\ }\href
  {\doibase 10.1126/science.aaf5541} {\bibfield  {journal} {\bibinfo  {journal}
  {Science}\ }\textbf {\bibinfo {volume} {354}},\ \bibinfo {pages} {1124}
  (\bibinfo {year} {2016})}\BibitemShut {NoStop}%
\bibitem [{\citenamefont {Dziom}\ \emph
  {et~al.}(2017{\natexlab{a}})\citenamefont {Dziom}, \citenamefont {Shuvaev},
  \citenamefont {Pimenov}, \citenamefont {Astakhov}, \citenamefont {Ames},
  \citenamefont {Bendias}, \citenamefont {B{\"o}ttcher}, \citenamefont
  {Tkachov}, \citenamefont {Hankiewicz}, \citenamefont {Br{\"u}ne},\ and\
  \citenamefont {Buhmann}}]{DziomQ2759}%
  \BibitemOpen
  \bibfield  {author} {\bibinfo {author} {\bibfnamefont {V.}~\bibnamefont
  {Dziom}}, \bibinfo {author} {\bibfnamefont {A.}~\bibnamefont {Shuvaev}},
  \bibinfo {author} {\bibfnamefont {A.}~\bibnamefont {Pimenov}}, \bibinfo
  {author} {\bibfnamefont {G.~V.}\ \bibnamefont {Astakhov}}, \bibinfo {author}
  {\bibfnamefont {C.}~\bibnamefont {Ames}}, \bibinfo {author} {\bibfnamefont
  {K.}~\bibnamefont {Bendias}}, \bibinfo {author} {\bibfnamefont
  {J.}~\bibnamefont {B{\"o}ttcher}}, \bibinfo {author} {\bibfnamefont
  {G.}~\bibnamefont {Tkachov}}, \bibinfo {author} {\bibfnamefont {E.~M.}\
  \bibnamefont {Hankiewicz}}, \bibinfo {author} {\bibfnamefont
  {C.}~\bibnamefont {Br{\"u}ne}}, \ and\ \bibinfo {author} {\bibfnamefont
  {L.~W.}\ \bibnamefont {Buhmann}, \bibfnamefont {H.~Molenkamp}},\ }\bibfield
  {title} {\enquote {\bibinfo {title} {Observation of the universal
  magnetoelectric effect in a {3D} topological insulator},}\ }\href
  {http://dx.doi.org/10.1038/ncomms15197} {\bibfield  {journal} {\bibinfo
  {journal} {Nat. Commun.}\ }\textbf {\bibinfo {volume} {8}},\ \bibinfo {pages}
  {15197} (\bibinfo {year} {2017}{\natexlab{a}})}\BibitemShut {NoStop}%
\bibitem [{\citenamefont {Tse}(2016)}]{TseNonlinear}%
  \BibitemOpen
  \bibfield  {author} {\bibinfo {author} {\bibfnamefont {Wang-Kong}\
  \bibnamefont {Tse}},\ }\bibfield  {title} {\enquote {\bibinfo {title}
  {Coherent magneto-optical effects in topological insulators: Excitation near
  the absorption edge},}\ }\href {\doibase 10.1103/PhysRevB.94.125430}
  {\bibfield  {journal} {\bibinfo  {journal} {Phys. Rev. B}\ }\textbf {\bibinfo
  {volume} {94}},\ \bibinfo {pages} {125430} (\bibinfo {year}
  {2016})}\BibitemShut {NoStop}%
\bibitem [{\citenamefont {Lee}\ and\ \citenamefont {Tse}(2017)}]{LeeNonlinear}%
  \BibitemOpen
  \bibfield  {author} {\bibinfo {author} {\bibfnamefont {W.-R.}\ \bibnamefont
  {Lee}}\ and\ \bibinfo {author} {\bibfnamefont {W.-K.}\ \bibnamefont {Tse}},\
  }\bibfield  {title} {\enquote {\bibinfo {title} {Dynamical quantum anomalous
  {H}all effect in strong optical fields},}\ }\href {\doibase
  10.1103/PhysRevB.95.201411} {\bibfield  {journal} {\bibinfo  {journal} {Phys.
  Rev. B}\ }\textbf {\bibinfo {volume} {95}},\ \bibinfo {pages} {201411}
  (\bibinfo {year} {2017})}\BibitemShut {NoStop}%
\bibitem [{\citenamefont {Ikebe}\ \emph {et~al.}(2010)\citenamefont {Ikebe},
  \citenamefont {Morimoto}, \citenamefont {Masutomi}, \citenamefont {Okamoto},
  \citenamefont {Aoki},\ and\ \citenamefont {Shimano}}]{IkebeGaAs}%
  \BibitemOpen
  \bibfield  {author} {\bibinfo {author} {\bibfnamefont {Y.}~\bibnamefont
  {Ikebe}}, \bibinfo {author} {\bibfnamefont {T.}~\bibnamefont {Morimoto}},
  \bibinfo {author} {\bibfnamefont {R.}~\bibnamefont {Masutomi}}, \bibinfo
  {author} {\bibfnamefont {T.}~\bibnamefont {Okamoto}}, \bibinfo {author}
  {\bibfnamefont {H.}~\bibnamefont {Aoki}}, \ and\ \bibinfo {author}
  {\bibfnamefont {R.}~\bibnamefont {Shimano}},\ }\bibfield  {title} {\enquote
  {\bibinfo {title} {Optical {H}all effect in the integer quantum {H}all
  regime},}\ }\href {\doibase 10.1103/PhysRevLett.104.256802} {\bibfield
  {journal} {\bibinfo  {journal} {Phys. Rev. Lett.}\ }\textbf {\bibinfo
  {volume} {104}},\ \bibinfo {pages} {256802} (\bibinfo {year}
  {2010})}\BibitemShut {NoStop}%
\bibitem [{\citenamefont {Stier}\ \emph {et~al.}(2015)\citenamefont {Stier},
  \citenamefont {Ellis}, \citenamefont {Kwon}, \citenamefont {Xing},
  \citenamefont {Zhang}, \citenamefont {Eason}, \citenamefont {Strasser},
  \citenamefont {Morimoto}, \citenamefont {Aoki}, \citenamefont {Zeng},
  \citenamefont {McCombe},\ and\ \citenamefont {Cerne}}]{StierGaAs}%
  \BibitemOpen
  \bibfield  {author} {\bibinfo {author} {\bibfnamefont {A.~V.}\ \bibnamefont
  {Stier}}, \bibinfo {author} {\bibfnamefont {C.~T.}\ \bibnamefont {Ellis}},
  \bibinfo {author} {\bibfnamefont {J.}~\bibnamefont {Kwon}}, \bibinfo {author}
  {\bibfnamefont {H.}~\bibnamefont {Xing}}, \bibinfo {author} {\bibfnamefont
  {H.}~\bibnamefont {Zhang}}, \bibinfo {author} {\bibfnamefont
  {D.}~\bibnamefont {Eason}}, \bibinfo {author} {\bibfnamefont
  {G.}~\bibnamefont {Strasser}}, \bibinfo {author} {\bibfnamefont
  {T.}~\bibnamefont {Morimoto}}, \bibinfo {author} {\bibfnamefont
  {H.}~\bibnamefont {Aoki}}, \bibinfo {author} {\bibfnamefont {H.}~\bibnamefont
  {Zeng}}, \bibinfo {author} {\bibfnamefont {B.~D.}\ \bibnamefont {McCombe}}, \
  and\ \bibinfo {author} {\bibfnamefont {J.}~\bibnamefont {Cerne}},\ }\bibfield
   {title} {\enquote {\bibinfo {title} {Terahertz dynamics of a topologically
  protected state: Quantum {H}all effect plateaus near the cyclotron resonance
  of a two-dimensional electron gas},}\ }\href {\doibase
  10.1103/PhysRevLett.115.247401} {\bibfield  {journal} {\bibinfo  {journal}
  {Phys. Rev. Lett.}\ }\textbf {\bibinfo {volume} {115}},\ \bibinfo {pages}
  {247401} (\bibinfo {year} {2015})}\BibitemShut {NoStop}%
\bibitem [{\citenamefont {Failla}\ \emph {et~al.}(2016)\citenamefont {Failla},
  \citenamefont {Keller}, \citenamefont {Scalari}, \citenamefont {Maissen},
  \citenamefont {Faist}, \citenamefont {Reichl}, \citenamefont {Wegscheider},
  \citenamefont {Newell}, \citenamefont {Leadley}, \citenamefont {Myronov},\
  and\ \citenamefont {Lloyd-Hughes}}]{FaillaGe}%
  \BibitemOpen
  \bibfield  {author} {\bibinfo {author} {\bibfnamefont {M.}~\bibnamefont
  {Failla}}, \bibinfo {author} {\bibfnamefont {J.}~\bibnamefont {Keller}},
  \bibinfo {author} {\bibfnamefont {G.}~\bibnamefont {Scalari}}, \bibinfo
  {author} {\bibfnamefont {C.}~\bibnamefont {Maissen}}, \bibinfo {author}
  {\bibfnamefont {J.}~\bibnamefont {Faist}}, \bibinfo {author} {\bibfnamefont
  {C.}~\bibnamefont {Reichl}}, \bibinfo {author} {\bibfnamefont
  {W.}~\bibnamefont {Wegscheider}}, \bibinfo {author} {\bibfnamefont {O.~J.}\
  \bibnamefont {Newell}}, \bibinfo {author} {\bibfnamefont {D.~R.}\
  \bibnamefont {Leadley}}, \bibinfo {author} {\bibfnamefont {M.}~\bibnamefont
  {Myronov}}, \ and\ \bibinfo {author} {\bibfnamefont {J.}~\bibnamefont
  {Lloyd-Hughes}},\ }\bibfield  {title} {\enquote {\bibinfo {title} {Terahertz
  quantum {H}all effect for spin-split heavy-hole gases in strained {G}e
  quantum wells},}\ }\href {http://stacks.iop.org/1367-2630/18/i=11/a=113036}
  {\bibfield  {journal} {\bibinfo  {journal} {New J. Phys.}\ }\textbf {\bibinfo
  {volume} {18}},\ \bibinfo {pages} {113036} (\bibinfo {year}
  {2016})}\BibitemShut {NoStop}%
\bibitem [{\citenamefont {Sebestyen}(1982)}]{SEBESTYEN_SSE_1982}%
  \BibitemOpen
  \bibfield  {author} {\bibinfo {author} {\bibfnamefont {T.}~\bibnamefont
  {Sebestyen}},\ }\bibfield  {title} {\enquote {\bibinfo {title} {Models for
  ohmic contacts on graded crystalline or amorphous heterojunctions},}\ }\href
  {\doibase https://doi.org/10.1016/0038-1101(82)90054-5} {\bibfield  {journal}
  {\bibinfo  {journal} {Solid-State Electron.}\ }\textbf {\bibinfo {volume}
  {25}},\ \bibinfo {pages} {543} (\bibinfo {year} {1982})}\BibitemShut
  {NoStop}%
\bibitem [{\citenamefont {Lakhani}(1984)}]{Lakhani_JAP_1984}%
  \BibitemOpen
  \bibfield  {author} {\bibinfo {author} {\bibfnamefont {A.~A.}\ \bibnamefont
  {Lakhani}},\ }\bibfield  {title} {\enquote {\bibinfo {title} {The role of
  compound formation and heteroepitaxy in indium-based ohmic contacts to
  {GaAs}},}\ }\href {\doibase 10.1063/1.334173} {\bibfield  {journal} {\bibinfo
   {journal} {J. Appl. Phys.}\ }\textbf {\bibinfo {volume} {56}},\ \bibinfo
  {pages} {1888} (\bibinfo {year} {1984})}\BibitemShut {NoStop}%
\bibitem [{\citenamefont {Baba}\ \emph {et~al.}(1983)\citenamefont {Baba},
  \citenamefont {Mizutani},\ and\ \citenamefont {Ogawa}}]{BabaJJAP83}%
  \BibitemOpen
  \bibfield  {author} {\bibinfo {author} {\bibfnamefont {T.}~\bibnamefont
  {Baba}}, \bibinfo {author} {\bibfnamefont {T.}~\bibnamefont {Mizutani}}, \
  and\ \bibinfo {author} {\bibfnamefont {M.}~\bibnamefont {Ogawa}},\ }\bibfield
   {title} {\enquote {\bibinfo {title} {Elimination of persistent
  photoconductivity and improvement in {Si} activation coefficient by {Al}
  spatial separation from {Ga} and {Si} in {Al}-{Ga}-{As}:{Si} solid system.
  {A} novel short period {AlAs}/n-{GaAs} superlattice},}\ }\href
  {http://stacks.iop.org/1347-4065/22/i=10A/a=L627} {\bibfield  {journal}
  {\bibinfo  {journal} {Jpn. J. Appl. Phys.}\ }\textbf {\bibinfo {volume}
  {22}},\ \bibinfo {pages} {L627} (\bibinfo {year} {1983})}\BibitemShut
  {NoStop}%
\bibitem [{\citenamefont {Nathan}(1986)}]{Nathan_SSE_1986}%
  \BibitemOpen
  \bibfield  {author} {\bibinfo {author} {\bibfnamefont {M.~I.}\ \bibnamefont
  {Nathan}},\ }\bibfield  {title} {\enquote {\bibinfo {title} {Persistent
  photoconductivity in {AlGaAs}/{GaAs} modulation doped layers and field effect
  transistors: A review},}\ }\href {\doibase
  https://doi.org/10.1016/0038-1101(86)90035-3} {\bibfield  {journal} {\bibinfo
   {journal} {Solid-State Electron.}\ }\textbf {\bibinfo {volume} {29}},\
  \bibinfo {pages} {167} (\bibinfo {year} {1986})}\BibitemShut {NoStop}%
\bibitem [{\citenamefont {Berreman}(1972)}]{Berreman}%
  \BibitemOpen
  \bibfield  {author} {\bibinfo {author} {\bibfnamefont {D.~W.}\ \bibnamefont
  {Berreman}},\ }\bibfield  {title} {\enquote {\bibinfo {title} {Optics in
  stratified and anisotropic media: 4x4-matrix formulation},}\ }\href {\doibase
  10.1364/JOSA.62.000502} {\bibfield  {journal} {\bibinfo  {journal} {J. Opt.
  Soc. Am.}\ }\textbf {\bibinfo {volume} {62}},\ \bibinfo {pages} {502}
  (\bibinfo {year} {1972})}\BibitemShut {NoStop}%
\bibitem [{\citenamefont {Dziom}\ \emph
  {et~al.}(2017{\natexlab{b}})\citenamefont {Dziom}, \citenamefont {Shuvaev},
  \citenamefont {Mikhailov},\ and\ \citenamefont {Pimenov}}]{DziomLight}%
  \BibitemOpen
  \bibfield  {author} {\bibinfo {author} {\bibfnamefont {V.}~\bibnamefont
  {Dziom}}, \bibinfo {author} {\bibfnamefont {A.}~\bibnamefont {Shuvaev}},
  \bibinfo {author} {\bibfnamefont {N.~N.}\ \bibnamefont {Mikhailov}}, \ and\
  \bibinfo {author} {\bibfnamefont {A.}~\bibnamefont {Pimenov}},\ }\bibfield
  {title} {\enquote {\bibinfo {title} {Terahertz properties of {D}irac fermions
  in $\mathrm{HgTe}$ films with optical doping},}\ }\href
  {http://stacks.iop.org/2053-1583/4/i=2/a=024005} {\bibfield  {journal}
  {\bibinfo  {journal} {2D Mater.}\ }\textbf {\bibinfo {volume} {4}},\ \bibinfo
  {pages} {024005} (\bibinfo {year} {2017}{\natexlab{b}})}\BibitemShut
  {NoStop}%
\bibitem [{\citenamefont {Shuvaev}\ \emph {et~al.}(2012)\citenamefont
  {Shuvaev}, \citenamefont {Astakhov}, \citenamefont {Br\"{u}ne}, \citenamefont
  {Buhmann}, \citenamefont {Molenkamp},\ and\ \citenamefont
  {Pimenov}}]{shuvaev_sst_2012}%
  \BibitemOpen
  \bibfield  {author} {\bibinfo {author} {\bibfnamefont {A.~M.}\ \bibnamefont
  {Shuvaev}}, \bibinfo {author} {\bibfnamefont {G.~V.}\ \bibnamefont
  {Astakhov}}, \bibinfo {author} {\bibfnamefont {C.}~\bibnamefont {Br\"{u}ne}},
  \bibinfo {author} {\bibfnamefont {H.}~\bibnamefont {Buhmann}}, \bibinfo
  {author} {\bibfnamefont {L.~W.}\ \bibnamefont {Molenkamp}}, \ and\ \bibinfo
  {author} {\bibfnamefont {A.}~\bibnamefont {Pimenov}},\ }\bibfield  {title}
  {\enquote {\bibinfo {title} {Terahertz magneto-optical spectroscopy in
  {$\mathrm{HgTe}$} thin films},}\ }\href
  {http://stacks.iop.org/0268-1242/27/i=12/a=124004} {\bibfield  {journal}
  {\bibinfo  {journal} {Semicond. Sci. Technol.}\ }\textbf {\bibinfo {volume}
  {27}},\ \bibinfo {pages} {124004} (\bibinfo {year} {2012})}\BibitemShut
  {NoStop}%
\bibitem [{\citenamefont {Palik}\ and\ \citenamefont {Furdyna}(1970)}]{Palik}%
  \BibitemOpen
  \bibfield  {author} {\bibinfo {author} {\bibfnamefont {E.~D.}\ \bibnamefont
  {Palik}}\ and\ \bibinfo {author} {\bibfnamefont {J.~K.}\ \bibnamefont
  {Furdyna}},\ }\bibfield  {title} {\enquote {\bibinfo {title} {Infrared and
  microwave magnetoplasma effects in semiconductors},}\ }\href
  {http://stacks.iop.org/0034-4885/33/i=3/a=307} {\bibfield  {journal}
  {\bibinfo  {journal} {Rep. Prog. Phys.}\ }\textbf {\bibinfo {volume} {33}},\
  \bibinfo {pages} {1193} (\bibinfo {year} {1970})}\BibitemShut {NoStop}%
\bibitem [{\citenamefont {Kuchar}\ \emph {et~al.}(1987)\citenamefont {Kuchar},
  \citenamefont {Meisels}, \citenamefont {Lim}, \citenamefont {Pichler},
  \citenamefont {Weimann},\ and\ \citenamefont {Schlapp}}]{KucharScripta}%
  \BibitemOpen
  \bibfield  {author} {\bibinfo {author} {\bibfnamefont {F.}~\bibnamefont
  {Kuchar}}, \bibinfo {author} {\bibfnamefont {R.}~\bibnamefont {Meisels}},
  \bibinfo {author} {\bibfnamefont {K.~Y.}\ \bibnamefont {Lim}}, \bibinfo
  {author} {\bibfnamefont {P.}~\bibnamefont {Pichler}}, \bibinfo {author}
  {\bibfnamefont {G.}~\bibnamefont {Weimann}}, \ and\ \bibinfo {author}
  {\bibfnamefont {W.}~\bibnamefont {Schlapp}},\ }\bibfield  {title} {\enquote
  {\bibinfo {title} {{H}all conductivity at microwave and submillimeter
  frequencies in the quantum {H}all effect regime},}\ }\href
  {http://stacks.iop.org/1402-4896/1987/i=T19A/a=013} {\bibfield  {journal}
  {\bibinfo  {journal} {Phys. Scr.}\ }\textbf {\bibinfo {volume} {1987}},\
  \bibinfo {pages} {79} (\bibinfo {year} {1987})}\BibitemShut {NoStop}%
\bibitem [{\citenamefont {Morimoto}\ \emph {et~al.}(2009)\citenamefont
  {Morimoto}, \citenamefont {Hatsugai},\ and\ \citenamefont
  {Aoki}}]{MorimotoPRL09}%
  \BibitemOpen
  \bibfield  {author} {\bibinfo {author} {\bibfnamefont {T.}~\bibnamefont
  {Morimoto}}, \bibinfo {author} {\bibfnamefont {Y.}~\bibnamefont {Hatsugai}},
  \ and\ \bibinfo {author} {\bibfnamefont {H.}~\bibnamefont {Aoki}},\
  }\bibfield  {title} {\enquote {\bibinfo {title} {Optical {H}all conductivity
  in ordinary and graphene quantum {H}all systems},}\ }\href {\doibase
  10.1103/PhysRevLett.103.116803} {\bibfield  {journal} {\bibinfo  {journal}
  {Phys. Rev. Lett.}\ }\textbf {\bibinfo {volume} {103}},\ \bibinfo {pages}
  {116803} (\bibinfo {year} {2009})}\BibitemShut {NoStop}%
\bibitem [{\citenamefont {Morimoto}\ \emph {et~al.}(2010)\citenamefont
  {Morimoto}, \citenamefont {Avishai},\ and\ \citenamefont
  {Aoki}}]{MorimotoPRB10}%
  \BibitemOpen
  \bibfield  {author} {\bibinfo {author} {\bibfnamefont {T.}~\bibnamefont
  {Morimoto}}, \bibinfo {author} {\bibfnamefont {Y.}~\bibnamefont {Avishai}}, \
  and\ \bibinfo {author} {\bibfnamefont {H.}~\bibnamefont {Aoki}},\ }\bibfield
  {title} {\enquote {\bibinfo {title} {Dynamical scaling analysis of the
  optical {H}all conductivity in the quantum {H}all regime},}\ }\href {\doibase
  10.1103/PhysRevB.82.081404} {\bibfield  {journal} {\bibinfo  {journal} {Phys.
  Rev. B}\ }\textbf {\bibinfo {volume} {82}},\ \bibinfo {pages} {081404}
  (\bibinfo {year} {2010})}\BibitemShut {NoStop}%
\bibitem [{\citenamefont {Lozovik}\ \emph {et~al.}(1984)\citenamefont
  {Lozovik}, \citenamefont {Farztdinov},\ and\ \citenamefont
  {Gevorkyan}}]{LozovikDelta}%
  \BibitemOpen
  \bibfield  {author} {\bibinfo {author} {\bibfnamefont {Yu.~E.}\ \bibnamefont
  {Lozovik}}, \bibinfo {author} {\bibfnamefont {V.~M.}\ \bibnamefont
  {Farztdinov}}, \ and\ \bibinfo {author} {\bibfnamefont {Zh.~S.}\ \bibnamefont
  {Gevorkyan}},\ }\bibfield  {title} {\enquote {\bibinfo {title} {Dynamic
  quantum {H}all effect},}\ }\href
  {http://www.jetpletters.ac.ru/ps/1287/article_19439.shtml} {\bibfield
  {journal} {\bibinfo  {journal} {JETP Lett.}\ }\textbf {\bibinfo {volume}
  {39}},\ \bibinfo {pages} {179} (\bibinfo {year} {1984})}\BibitemShut
  {NoStop}%
\bibitem [{\citenamefont {Apenko}\ and\ \citenamefont
  {Lozovik}(1985)}]{ApenkoSmooth}%
  \BibitemOpen
  \bibfield  {author} {\bibinfo {author} {\bibfnamefont {S.~M.}\ \bibnamefont
  {Apenko}}\ and\ \bibinfo {author} {\bibfnamefont {Yu.~E.}\ \bibnamefont
  {Lozovik}},\ }\bibfield  {title} {\enquote {\bibinfo {title} {Quantization of
  the {H}all conductivity of a two-dimensional electron gas in a strong
  magnetic field},}\ }\href
  {http://www.jetp.ac.ru/cgi-bin/r/index/e/62/2/p328?a=list} {\bibfield
  {journal} {\bibinfo  {journal} {J. Exp. Theor. Phys.}\ }\textbf {\bibinfo
  {volume} {62}},\ \bibinfo {pages} {328} (\bibinfo {year} {1985})}\BibitemShut
  {NoStop}%
\bibitem [{\citenamefont {Prange}(1981)}]{PrangeDelta}%
  \BibitemOpen
  \bibfield  {author} {\bibinfo {author} {\bibfnamefont {R.~E.}\ \bibnamefont
  {Prange}},\ }\bibfield  {title} {\enquote {\bibinfo {title} {Quantized {H}all
  resistance and the measurement of the fine-structure constant},}\ }\href
  {\doibase 10.1103/PhysRevB.23.4802} {\bibfield  {journal} {\bibinfo
  {journal} {Phys. Rev. B}\ }\textbf {\bibinfo {volume} {23}},\ \bibinfo
  {pages} {4802} (\bibinfo {year} {1981})}\BibitemShut {NoStop}%
\bibitem [{\citenamefont {Lozovik}(2017)}]{LozovikPrivateCommunications}%
  \BibitemOpen
  \bibfield  {author} {\bibinfo {author} {\bibfnamefont {Yu.~E.}\ \bibnamefont
  {Lozovik}},\ }\href@noop {} {\enquote {\bibinfo {title} {private
  communications},}\ } (\bibinfo {year} {2017})\BibitemShut {NoStop}%
\bibitem [{\citenamefont {Tsuneya}\ and\ \citenamefont
  {Yasutada}(1974)}]{AndoJPs}%
  \BibitemOpen
  \bibfield  {author} {\bibinfo {author} {\bibfnamefont {A.}~\bibnamefont
  {Tsuneya}}\ and\ \bibinfo {author} {\bibfnamefont {U.}~\bibnamefont
  {Yasutada}},\ }\bibfield  {title} {\enquote {\bibinfo {title} {Theory of
  oscillatory g factor in an {MOS} inversion layer under strong magnetic
  fields},}\ }\href {\doibase 10.1143/JPSJ.37.1044} {\bibfield  {journal}
  {\bibinfo  {journal} {J. Phys. Soc. Jpn.}\ }\textbf {\bibinfo {volume}
  {37}},\ \bibinfo {pages} {1044} (\bibinfo {year} {1974})}\BibitemShut
  {NoStop}%
\bibitem [{\citenamefont {Shuvaev}\ \emph {et~al.}(2016)\citenamefont
  {Shuvaev}, \citenamefont {Dziom}, \citenamefont {Kvon}, \citenamefont
  {Mikhailov},\ and\ \citenamefont {Pimenov}}]{ShuvaevUniversal}%
  \BibitemOpen
  \bibfield  {author} {\bibinfo {author} {\bibfnamefont {A.}~\bibnamefont
  {Shuvaev}}, \bibinfo {author} {\bibfnamefont {V.}~\bibnamefont {Dziom}},
  \bibinfo {author} {\bibfnamefont {Z.~D.}\ \bibnamefont {Kvon}}, \bibinfo
  {author} {\bibfnamefont {N.~N.}\ \bibnamefont {Mikhailov}}, \ and\ \bibinfo
  {author} {\bibfnamefont {A.}~\bibnamefont {Pimenov}},\ }\bibfield  {title}
  {\enquote {\bibinfo {title} {Universal {F}araday rotation in $\mathrm{HgTe}$
  wells with critical thickness},}\ }\href {\doibase
  10.1103/PhysRevLett.117.117401} {\bibfield  {journal} {\bibinfo  {journal}
  {Phys. Rev. Lett.}\ }\textbf {\bibinfo {volume} {117}},\ \bibinfo {pages}
  {117401} (\bibinfo {year} {2016})}\BibitemShut {NoStop}%
\bibitem [{\citenamefont {Iordansky}(1982)}]{IordanskySmooth}%
  \BibitemOpen
  \bibfield  {author} {\bibinfo {author} {\bibfnamefont {S.V.}\ \bibnamefont
  {Iordansky}},\ }\bibfield  {title} {\enquote {\bibinfo {title} {On the
  conductivity of two dimensional electrons in a strong magnetic field},}\
  }\href {\doibase 10.1016/0038-1098(82)91141-3} {\bibfield  {journal}
  {\bibinfo  {journal} {Solid State Commun.}\ }\textbf {\bibinfo {volume}
  {43}},\ \bibinfo {pages} {1 -- 3} (\bibinfo {year} {1982})}\BibitemShut
  {NoStop}%
\bibitem [{\citenamefont {Kazarinov}\ and\ \citenamefont
  {Luryi}(1982)}]{KazarinovSmooth}%
  \BibitemOpen
  \bibfield  {author} {\bibinfo {author} {\bibfnamefont {R.~F.}\ \bibnamefont
  {Kazarinov}}\ and\ \bibinfo {author} {\bibfnamefont {S.}~\bibnamefont
  {Luryi}},\ }\bibfield  {title} {\enquote {\bibinfo {title} {Quantum
  percolation and quantization of {H}all resistance in two-dimensional electron
  gas},}\ }\href {\doibase 10.1103/PhysRevB.25.7626} {\bibfield  {journal}
  {\bibinfo  {journal} {Phys. Rev. B}\ }\textbf {\bibinfo {volume} {25}},\
  \bibinfo {pages} {7626--7630} (\bibinfo {year} {1982})}\BibitemShut {NoStop}%
\end{thebibliography}%
\end{document}